# Give Me a Like: How HIV/AIDS Nonprofit Organizations Can Engage Their Audience on Facebook

Yu-Chao Huang, Yi-Pin Lin, Gregory D. Saxton

Yu-Chao Huang, Ph.D., is affiliated with Department of Indigenous Languages and Communication, National Dong Hwa University, Taiwan.
Yi-Pin Lin, M.A., and Gregory D. Saxton, Ph.D., are affiliated with Department of Communication, University at Buffalo, SUNY.

Address correspondence to Yu-Chao Huang, 1, Sec. 2, Da Hsueh Rd., Hualien 97401, Taiwan.
E-mail: yc@mail.ndhu.edu.tw
Tel: + 886-3863-5831
Fax: +886-3863-5830

## Acknowledgements

The authors would like to sincerely thank Dr. Thomas Feeley, Weiai Wayne Xu, Jason Grenier, and two anonymous reviewers for helpful comments and suggestions.

# Give Me a Like: How HIV/AIDS Nonprofit Organizations Can Engage Their Audience on Facebook

**ABSTRACT**

With the rapid proliferation and adoption of social media among healthcare professionals and organizations, social media-based HIV/AIDS intervention programs have become increasingly popular. However, the question of the effectiveness of the HIV/AIDS messages disseminated via social media has received scant attention in the literature. The current study applies content analysis to examine the relationship between Facebook messaging strategies employed by 110 HIV/AIDS nonprofit organizations and audience reactions in the form of liking, commenting, and sharing behavior. The results reveal that HIV/AIDS nonprofit organizations often use informational messages as one-way communication with their audience instead of dialogic interactions. Some specific types of messages, such as medication-focused messages, engender better audience engagement; in contrast, event-related messages and call-to-action messages appear to translate into lower corresponding audience reactions. The findings provide guidance to HIV/AIDS organizations in developing effective social media communication strategies.

**Keywords**: audience engagement, AIDS, Facebook, health messages, social media

# INTRODUCTION

Social media has become a primary channel for health professionals to connect to their peers, patients, and the community at large (Anikeeva & Bywood, 2013; Antheunis, Tates, & Nieboer, 2013; Harris, Choucair, Maier, Jolani, & Bernhardt, 2014). Governmental sectors as well as organizations are applying new social technologies to reach out to the public. For example, the Centers for Disease Control and Prevention (CDC) endeavors to connect to the general public via numerous social networking sites (SNS) such as Facebook, Twitter, and YouTube and has even published guidelines for composing effective messages on social media (CDC, 2012). The interactive communication channels and networked environment provided by social media are widely taken up by organizations for health promotion (Ramanadhan, Mendes, Rao, & Viswanath, 2013), as evidenced by the innovative application of social media in initiating HIV/AIDS prevention interventions (e.g., Jaganath, Gill, Cohen & Young, 2012; Ramallo et al., 2015, Rhodes, et al., 2014).

While social media has great potential and offers multiple benefits for communication aimed at promoting health, some health professionals are still not yet equipped to exploit these features, and some even refuse to adopt social media due to a lack of relevant training or knowledge (Antheunis et al., 2013; Chou, Hunt, Beckjord, Moser, & Hesse, 2009; Usher, 2012). Since social media is not limited in terms of its ability to reach out to the population, regardless of location, it is particularly well-suited for community-based HIV prevention (for a review, see Young & Jaganath, 2013). HIV/AIDS related organizations should thus equip themselves with social media skills that will allow them to extend their offline resources to online venues, so as to be able to provide information and remote assistance for those who are in need. However, what makes for relevant and engaging content in the social media messages issued by HIV/AIDS-related organizations has received scant attention in the literature. Thus, the current study aims to investigate how HIV/AIDS-related organizations employ social media to achieve their communication goals. A message-level analysis forms the core of the current study, in which three variables are examined: communicative functions, mission relevance and focus, and use of hashtags. Each of these variables will be discussed below.

# COMMUNICATIVE FUNCTIONS

Lovejoy and Saxton (2012) analyzed 2,437 tweets produced by 100 large US nonprofit organizations and established a framework that identified three dimensions of messaging strategies: *information*, *community*, and *action*. This became known as the Information-Community-Action (ICA) framework. *Informational* messages denote one-way communication, including organizational announcements, facts, event information and updates. *Community* messages are critical in forming bonds with organizational followers. These messages express appreciation, give acknowledgement to causes, or attempt to elicit feedback or discussion. Lastly, *Action* messages mainly appeal to the audience to do something to further organizational objectives, such as making donations, participating in events, lobbying (e.g., asking the audience to advocate a specific policy), or helping disseminate messages. Lovejoy and Saxton's (2012) study found that nonprofit organizations had a relatively low likelihood of utilizing the interactive nature of social media for two-way communication. Most tweets were informative (58.6%), whereas community- (25.8%) and action-oriented (15.6%) tweets were posted much less often.

Later studies observed audience responses to messages disseminated by nonprofit organizations and provided a more comprehensive structure of not only how practitioners composed messages, but also how stakeholders reacted to the messages. For example, Saxton and Waters (2014) found that community-oriented messages attracted more likes and comments than informational messages, while informational messages generated more shares, on average, than the other two categories. In order to determine how HIV/AIDS-related organizations communicate with their audience through social media, the ICA framework is employed in the current study to examine the proportion of messages comprising each communicative function among the messages they produce, and the association between the message content and audience engagement in the form of liking, commenting, and sharing.

## MISSION RELEVANCE AND FOCUS

The concept of *mission relevance* was introduced by Guo and Saxton (2014) in order to analyze the tweets posted by nonprofit advocacy organizations. In their study, messages aimed at explicating organizational advocacy were classified as mission-related or "strategic" messages, whereas those that were not relevant to their advocacy mission were categorized as non-mission related or "support" messages. Following this scheme, Covert et al. (In press) applied the mission-relevance framework to examine audience responsiveness to the Facebook wall postings produced by organ procurement organizations (OPOs) and found that mission-related messages were associated with higher audience engagement. The objectives of HIV/AIDS-related organizations are mostly in line with the CDC's emphasis on intervention (see http://www.cdc.gov/hiv/dhap/progress); these organizations strive either to facilitate prevention or to provide a better quality of life for patients. In all cases, organizations' capacity to fulfill their mission is considered essential. Drawing upon these previous two studies and the objectives of HIV/AIDS-related organizations, the three major types of mission foci identified in the current study are *prevention*, *patient advocacy*, and *capacity building*. In contrast, purely "social" messages and off-topic messages are considered non-mission related.

## HASHTAGS

Another communication tool that originated on Twitter but can now be seen on most social media platforms is the hashtag. Hashtags allow users to highlight the topic of tweets or posts. A hashtag usually refers to a specific topic (e.g., #HIV) or shortened form of expression (e.g., #FF: Follow Friday), and it makes the search for content related to a certain topic easier. Hashtags were used in 30% of Twitter posts in 2012 (Lovejoy et al., 2012), with this number increasing to 60.5% in 2014 (Guo & Saxton, 2014). The Lovejoy et al. and Guo & Saxton studies extracted tweets from 73 and 188 nonprofit organizations, respectively, and both indicated a high rate of adoption of the hashtag function.

Facebook is another social media giant. After years of operation, Facebook only launched its hashtag function in 2013 (Facebook, 2013). Because this function has only been available for the past three years, the actual adoption rates of the hashtag function on Facebook have not been widely explored. The use of hashtags can reveal the aggregate information about a meme, topic, event, or social movement. Indeed, in some cases, it has helped foster the impact of movements. For instance, during the Arab Spring, the hashtags #egypt and #libya were widely used and disseminated, which affected the interaction and message flow among Twitter users and, in turn, increased the impact of the movement (Bruns, Highfield, & Burgess, 2013).

In sum, the purpose of this study is to extend a previously established framework to gauge the efficacy of Facebook posts from HIV/AIDS nonprofit organizations. At the same time, by analyzing these messages, the current study also aims to identify messaging strategies utilized by this specific type of organization. In the field of health promotion, Covert et al. (In press) have conducted a preliminary study to apply the ICA framework in the context of organ donation, which provides a solid foundation for future message-level, practitioner-oriented analyses in the field of health communication. This study, by applying and extending their framework, hopes to lay the groundwork that could assist HIV/AIDS organizations in developing effective social media communication strategies.

**METHOD**

SAMPLE

To obtain a list of HIV/AIDS nonprofit organizations that qualify as tax-exempt organizations, the authors requested a list of charitable organizations from the National Center for Charitable Statistics (NCCS) database, which includes all non-religious US charities with revenues greater than $25,000. From this database all organizations with the National Taxonomy Exempt Entities (NTEE) code G81 (*AIDS*) were initially selected. There are thousands of organizations under this category, and thus to narrow down the list, only those with total revenue of over 1 million U.S. dollars were selected as our sample, yielding a total of 171 organizations. Each of the organizations was checked to ensure that the organization is actually HIV/AIDS-oriented, given that a few were misclassified in the NCCS data. In determining whether the organization is HIV/AIDS-related, names and organizational objectives were used to make the judgments. Notably, if the name of the organization contains "HIV" or "AIDS," then it was included in the sample. In cases where there was no mention of HIV/AIDS in their names, the organizations' statements of their missions or objectives were used as the selection criterion. Of the 171 organizations initially considered, 35 were determined not to be HIV/AIDS-focused, while 19 did not have Facebook pages and 7 did not allow public access to their Facebook IDs. In sum, our sample consisted of 110 organizations, a list of which is shown in Table 1.

PLACE TABLE 1 ABOUT HERE

DATA COLLECTION

To make sure that the sample messages fully represent organizations' messaging strategies, the time frame for sampling is an entire year (2014). The messages were retrieved by accessing a Facebook application-programming interface (API) via a Python script specifically written to download the 110 organizations' posts. A total of 23,601 messages was retrieved from these organizations in 2014, and of these messages 1,500 were randomly selected for manual coding.

PROCEDURES

The general guidelines for the coding scheme were first developed by Lovejoy and Saxton (2012) and subsequently modified by Covert et al. (In press) for health promotion organizations. In order to make the codebook suitable for the current study, two of the co-authors independently coded the first 150 Facebook statuses and identified the sub-codes under each communicative function. Coding discrepancies were discussed and resolved. In the meantime, the codebook was refined to reach agreement. To make sure the coding framework encompassed all elements and was able to reach a satisfactory level of inter-coder reliability, the two coders continued to code the following 100 statuses. After three iterations of coding sets of 100 posts and refining the coding rules, the inter-coder reliability for each dimension reached acceptable levels of agreement. In the final round of coding, Cohen's kappa results for mission relevance reached .94, for mission focus reached .96, for communicative function (ICA) reached .96, and for ICA sub-codes reached .95. With such high levels of inter-coder agreement, the remaining 1050 statuses were divided and coded independently by the two coders.

## MESSAGE CODING

*Communicative Function*. The communicative function comprises three dimensions: information, community, and action. Informational posts are unidirectional, meaning the message is disseminated by the organization to spread facts and knowledge about HIV/AIDS, event information, organizational announcements, etc. Community-based posts, in contrast, are those that attempt to prompt audience responses and to encourage interactions with fellow organizations or its audience. Action-oriented messages call the audience to action for causes, events, or for the organization itself (e.g., soliciting donations).

*Mission Relevance & Focus*. The current study refines the concept of mission relevance (Covert et al., In press; Guo & Saxton, 2014) by, in addition to determining simply whether the message is mission-relevant, further categorizing mission-relevant messages into those aimed at, respectively, prevention, patient advocacy, and capacity building. Prevention messages are related to preventive measures or practices that are designed to constrain the spread of HIV. Patient advocacy information is intended to improve the quality of life of people living with HIV. Capacity building refers to organizations' ability to successfully carry out organizational missions; for health promotion organizations in particular, building capacity involves effectively maximizing the health effect of their programs, including tangible dimensions such as volunteers and donations and intangible dimensions such as skills and strategies (De Vita, 2001; Hawe, Noort, King, & Jordens, 1997; Kaplan, 2000). However, the current study does not aim at further breaking down tactics aimed at building capacity, so whenever a post refers to either tangible or intangible resources, it is categorized under "capacity building." A list of sub-categories (along with their definitions and examples) of each mission and communicative function is displayed in Appendix 1.

*Control Variables*. In addition to the variables that are of principal interest to the current study (i.e., mission relevance and focus, ICA, and hashtag count), two other factors that were regarded as possible contributors to audience engagement were included as control variables. In particular, two organizational-level factors, the number of Facebook followers and organizational assets (according to their most recent IRS Form 990) were included as controls.

*Dependent Variables: Audience Engagement*. On Facebook, the number of likes, comments, and shares can measure audience engagement (Saxton & Waters, 2014). Liking suggests the post is appreciated by the audience; commenting is a way of responding to a post and developing dialogue with other followers and organizations; lastly, sharing allows users to actually disseminate the information to their networks, which would then increase the exposure of the shared content among a wider audience.

## DATA ANALYSIS

The dependent variable of this study is audience engagement, which is represented by the number of likes, comments, and shares, and thus is a ratio-level count variable. A negative binomial regression analysis is suitable for this study because these variables are over-dispersed, which means their variance exceeds the mean, making this type of regression appropriate for the analysis. Two main sets of binomial regressions are presented. In the first set we include variables to tap the effects of hashtags, the three main mission foci, and the three primary I-C-A categories. In the second set we replace the three I-C-A categories with an expanded set of information, community, and action sub-codes. For each set we run regressions with the number of likes, comments and shares, respectively, as the dependent variable, making for 6 total regressions.

**RESULTS**

DESCRIPTIVE ANALYSES

The frequencies for each message type, as well as the descriptive statistics associated with audience engagement, are shown in Table 2. Examples of the different types of messages in each category, meanwhile, are shown in Appendix 1. Results show that nearly half of the 1,500 posts were informational (46.20%). Information-based posts flow one-way, from the organization to the audience, and their primary function is to inform and to disseminate materials that are of interest to the audience. Informational messages consist of various categories, and the most commonly seen were *HIV/AIDS info/news* (22.22% of informational posts) and *Organizational news/announcement* (20.63%), followed by *Event info* (16.74%) and *Event update* (13.28%). *HIV/AIDS info/news* messages disseminate knowledge or facts about the disease and those communities which are considered high-risk, such as lesbian, gay, bisexual, and transgender (LGBT) communities. This type of post engendered a moderate level of audience response compared to other types.

PLACE TABLE 2 ABOUT HERE

In terms of the number of likes, *Event info* ($M$ = 4.18, $SD$ = 6.05), which provided details of an event, such as time, place, ticketing information, or a direct link to the event, elicited significantly less response than *Event update* ($M$ = 21.85, $SD$ = 65.84). In comparison to the message types that were more frequently seen, *Medication*, which only accounted for 3.61% of informational messages, generated the highest number of likes ($M$ = 107.24, $SD$ = 225.24), comments ($M$ = 4.72, $SD$ = 14.42), and shares ($M$ = 26.16, $SD$ = 55.07) on average, compared to other categories. However, *Complementary support* (5.91%), which is related to patients' welfare and benefits in much the same was as *Medication*, did not generate a similar audience reaction. In addition, despite making up only 5.62% of the informational messages, *Awareness*, which chiefly served as reminders of preventive measures, received greater attention than *HIV/AIDS info/news* when it came to all three kinds of audience response.

In terms of community-based messages, the primary purpose is to maintain a relationship and build a connection to other organizations and the targeted audience. These messages focus on prompting interactivity. Among 266 community-based messages, *Recognition* accounted for 74.43% of all community-based messages, followed by *AIDS-related day* (11.65%), *Dialogue* (7.52%), and *National holiday/Holiday* (6.39%). Nonetheless, *AIDS-related day* generated the highest number of likes ($M$ = 34.65, $SD$ = 97.59) and shares ($M$ = 8.74, $SD$ = 28.52), and *Dialogue*, which is considered to be associated with higher levels of interactivity and prompting of a direct response, only elicited a slightly higher number of comments ($M$ = 0.90, $SD$ = 1.41) than other messages. *National holiday/Holiday* generated the least number of all three types of audience reactions, reflecting the fact that it is the least related to HIV/AIDS.

Lastly, action-oriented messages are intended to mobilize the audience to actually do something for the organizations, such as making a donation, participating in or registering for an event, reading an article, or sharing something within their own social media network (see Appendix 1 for specific examples). A large portion of the messages were categorized as *Event promotion* (51.02% of the action-oriented messages); these posts usually included verbs (e.g. "Participate in," "Register for") to appeal to audience members to attend. Apart from event promotion, *Get tested* (14.97%) and *Donation* (15.71%) messages prompted more likes than the others. In particular, *Donation* ($M = 145.44$, $SD = 1192.77$) messages generated the most likes on average, which overwhelmingly exceeded all other types of messages. One thing worth noting is that for *Media action* ($M = .33$, $SD = .49$), which comprised posts that specifically requested social media action (e.g. share, like, follow, and comment), these posts had the least number of shares. In contrast, *Viewing action* ($M = 2.55$, $SD = 7.81$) generated the most shares compared to other action-oriented messages.

## MULTIVARIATE ANALYSES

The results of the six negative binomial regression analysis are displayed in Table 3. All three models contain the same set of hashtag and mission focus and control variables. What varies across models is the specificity of the ICA variables and the particular dependent variable examined. The first three models test the effects of the broad information-community-action framework as independent variables with likes, comments and shares, respectively, as the dependent variable, while the last three models break the ICA framework out into multiple sub-codes for each category.

## PLACE TABLE 3 ABOUT HERE

In regards to the association with the number of likes ($\chi^2 = 14661$, $p < .01$), two mission foci, *Prevention* ($\beta = .73$, $p < .05$) and *Capacity building* ($\beta = .73$, $p < .01$), predicted more likes as opposed to non-mission related posts. As for the ICA framework, the omitted (comparison, or baseline) category is informational messages. In comparison to informational messages, the negative coefficient on *Action* indicates that action-oriented messages ($\beta = -.60$, $p < .01$) received significantly fewer likes, while the non-significant coefficient on *Community* indicates that community-based messages are not significantly different from informational messages in terms of the number of likes generated. The incorporation of hashtags surprisingly did not help obtain a higher volume of reactions; in fact, the presence of hashtags turned out to generate fewer likes. In terms of the organizational control variables, the number of followers was found not to be positively or negatively associated with the audience's liking behavior, yet organizations with larger assets ($\beta = .00$, $p < .01$) generated more likes.

The fourth model in Table 3, meanwhile, shows the association with liking behaviors for the various sub-codes of the three information-community-action communicative functions ($\chi^2 = 8602.1$, $p < .01$). With *Other* as the (omitted) baseline indicator, the following indicators were found to generate a significantly higher or lower number of likes. Among informational messages, those that were related to *Organizational news/announcement* ($\beta = .69$, $p < .05$) and *Medication* ($\beta = 1.11$, $p < .05$) generated a higher number of likes, yet *Event info* ($\beta = -1.15$, $p < .01$) engendered a decreased number of likes. Notably, the four indicators of community-based messages did not result in significantly more or less likes when compared to *Other* messages. In terms of action-based messages, messages that were calling for the audience to get an HIV test ($\beta = -1.16$, $p < .05$) obtained less likes, while the rest of the action sub-code indicators did not obtain significant coefficients.

Model 2, focusing on commenting behavior, was significant as well ($\chi^2$ = 19100, $p < .01$). In particular, posts related to *Prevention* ($\beta$ = 1.27, $p < .01$), *Patient advocacy* ($\beta$ = 1.02, $p < .01$), and *Capacity building* ($\beta$ = .87, $p < .01$), compared to non-mission posts, prompted more comments. However, action-oriented messages ($\beta$ = -.80, $p < .01$), compared to informational messages, received significantly fewer comments. Also, the inclusion of hashtags once again was not associated with the number of comments. In regards to the indicators of the three sub-codes of information, community, and action messages shown in model 5, most of the indicators did not engender obvious increases or decreases in the number of comments. However, three types of messages in particular resulted in a decline in the number of comments. The posting of *Complementary support* ($\beta$ = -1.49, $p < .01$), *Event info* ($\beta$ = -2.2, $p < .01$), and *Event update* ($\beta$ = -1.87, $p < .05$) were associated with significantly fewer comments, and among action-oriented messages, posts regarding *Get tested* ($\beta$ = -1.73, $p < .05$) and *Event promotion* ($\beta$ = -1.13, $p < .05$), which were most often employed by the organizations, turned out to lead to a decrease in the number of comments.

Finally, in terms of shares ($\chi^2$ =5981.9, $p < .01$) shown in model 3, the association of the independent and control variables with audience engagement was in accordance with the results pertaining to liking behavior. The mission posts focusing on *Prevention* ($\beta$ = .9, $p < .05$) were shown to be effective predictors of sharing behavior. Compared to informational posts, action-based ($\beta$ = -.78, $p < .01$) posts did not positively predict sharing response; instead, followers shared these posts less often. In general, the results suggested that action-based messages were not able to effectively engender sharing behavior, as opposed to information-oriented and community-based messages. In terms of the control variables, in turn, the number of followers and the size of the organization were not related to the number of shares. Lastly, with respect to the relationship between the number of shares and the various information-community-action sub-codes shown in the sixth model ($\chi^2$ = 4701.9, $p < .01$), the findings indicated that *Event info* ($\beta$ = -1.48, $p < .01$), *AIDS-related day* ($\beta$ = -1.71, $p < .1$), *Get tested* ($\beta$ = -2.4, $p < .1$), *Event promotion* ($\beta$ = -1.1, $p < .01$), and *Donation* ($\beta$ = -1.21, $p < .1$) messages generated notably fewer shares.

To provide some context to the data we present in Figure 1 a graphical summary of some key results. Rather than present all findings we focus on the first model from Table 3, which provides a good overview of the main findings. Figure 1 shows the number of likes a message with various conditions is predicted to obtain based on the coefficients in model 1. Specifically, the figure shows nine data points, with each data point representing the number of likes predicted by our model. The first three data points show the predicted number of likes for *Information* (31 likes)*, Community* (37 likes)*,* and *Action* messages (17 likes), respectively, for non-mission-related messages sent by organizations with mean values of assets and number of followers. The next three data points indicate the predicted number of likes for the three types of mission-oriented informational messages, namely, *Prevention* (64 likes), *Patient Advocacy* (55 likes), and *Capacity-Building* (65 likes). The final three data points, meanwhile, show the predicted number of likes for messages that combine a *Community* audience orientation with a *Prevention* (76 likes), *Patient Advocacy* (64 likes), and *Capacity-Building* (76 likes) mission focus. This figure effectively shows there are substantial differences in the expected number of likes that can be expected depending on the type of audience orientation taken, with action-oriented non-mission-related messages receiving half as many predicted likes as community-oriented non-mission-related messages. The figure also shows that all three types of mission-related Facebook posts are predicted to receive a much larger number of audience likes than non-mission-related posts.

Finally, the figure shows that messages that combine a mission orientation with the right audience focus are expected to engender the highest yield in audience engagement. Similar findings obtain with respect to the number of comments and shares.

PLACE FIGURE 1 ABOUT HERE

**DISCUSSION AND CONCLUSIONS**
The results of the study revealed how HIV/AIDS nonprofit organizations utilize Facebook and shed light on the extent to which their posts were able to effectively elicit audience reactions in the form of liking, commenting, and sharing. First, this study discovered that information-based (46.2%) messages were the most commonly employed type among these organizations, followed by action-oriented (36.1%) and community-oriented messages (17.7%). The proportion of each communicative function was somewhat inconsistent with those of Covert et al. (In press), in which each type of communicative function was more or less equally adopted by organ procurement organizations. Among information-oriented posts, 22% were *HIV/AIDS information or news*, while the second and third most commonly seen messages were *Organizational news/announcement* (20.63%) and *Event update* (13.28%), which greatly outnumbered those that were more directly pertinent to HIV/AIDS patients (i.e., *Medication* and *Complementary support*). In terms of interaction with the target audience, 13.2% of the total 1500 posts were showing appreciation to other organizations, significant individuals, or volunteers. Additionally, these organizations frequently advertised events via Facebook (51.02% of action-based posts, and 18.40% of total posts) by sending messages to encourage audience engagement, suggesting they spent a considerable amount of effort broadcasting their events and soliciting participation. However, as seen in Table 3 the volume of action-oriented messages did not translate into corresponding levels of audience reaction. For instance, posts making strong appeals for the audience to undergo HIV testing (e.g., "get tested" and "know your status") yielded significantly fewer reactions than basic informational messages. Likewise, messages that were calling for attendance at events and requesting donations received notably fewer shares, which is consistent with Ramanadhan et al. (2013), who also found that messages attempting to solicit donations generated fewer responses. Of the information-oriented messages, medication-related information, although only accounting for 1.67% of the 1,500 posts, received the highest number of likes, comments, and shares on average among all three communicative functions, implying that medication-related topics such as Pre-exposure prophylaxis (PrEP) may have better connections with audiences.

The disproportionate employment of various messaging strategies reflected organizations' focus and expectations in relation to social media. The majority of informational and action-based messages did not lead to higher engagement, suggesting that the organizations posting them have not yet mastered the use of social media as a facilitator of communication and interaction. The results from the negative binomial regressions are illustrative for these organizations in terms of *how* and *where* they should adjust their messaging strategies. It is crucial to note that certain messages, such as action-oriented messages (e.g. *Event promotion*), as well as event-related messages (e.g. *Event info*), predictably generated less audience reaction, while messages that were meant to initiate communication (i.e., *Dialogue*) were not able to generate notable amounts of feedback.

Another focus of this study was to see if audiences responded differently depending on each mission focus. This study classified mission-related messages into three major categories:

prevention, patient advocacy, and capacity building. Of the 1,500 posts examined, 63% were related to building capacity, implying that the organizations put considerable effort into building relationships via various events and activities designed to facilitate the organizations meeting their objectives, such as requesting donations or expressing appreciation to individuals for their contributions to the organizations. Table 3 indicated that mission-related messages were effective in generating a higher number of each type of audience response, and also had higher odds of eliciting reactions, as opposed to non-mission related messages. With respect to the quantity of reactions relating to each focus, HIV/AIDS prevention messages received the most attention from audiences.

Lastly, hashtags represent a relatively new function on Facebook (as of 2013), and their influence is yet to be fully explored. The study's organizations, on average, included only 0.48 hashtags in each post. According to Twitter, a maximum of two hashtags is the appropriate quantity to effectively convey messages without hindering reading or confusing readers ("How to use hashtags," n.d., para 7.). Considering that Facebook does not have any limitation in terms of the number of characters, *how* posts are composed and where hashtags should be inserted needs to be carefully assessed.

This study is among the first to observe how HIV/AIDS nonprofit organizations utilize Facebook to establish a connection with their followers, and has identified these organizations' unique messaging strategies that distinguish them from general nonprofit organizations or organ procurement organizations. The indicators of each communicative function and the corresponding results could serve as a preliminary guideline for these organizations. HIV/AIDS nonprofit organizations should reconsider their messaging strategies and know which types of posts are of the most interest to their followers. Since highly utilized message types do not necessarily translate into corresponding audience reactions, the organizations could take the opportunity to reevaluate their social media approaches and tailor messages in a way that will better allow them to elicit the expected responses. The frequency and mix of mission-related messages should conform to strategically designed goals.

**LIMITATION AND FUTURE RESEARCH**

One of the limitations of the current study is that the composition of the organizations' followers is unknown. As Covert et al. (In press) addressed, organizations should know the traits of their followers and their intentions in relation to following the organizations. Some may be people living with HIV (PLWH), some are family members, and others may belong to fellow organizations. Knowing the composition of followers should provide answers to why certain types of messages are not notably favored by the audience. The findings will be important for these HIV/AIDS specific organizations, since from there, these organizations can further develop and customize their messages for their targeted audience to generate desired responses.

Additionally, future research could delve further into message categories. A few sub-codes were collapsed into other overarching categories, such as lobbying and LGBT-related information. Although relatively scarce, these messages sometimes make strong appeals to mobilize their followers in an attempt to have an impact on public issues or strive for improvement in the welfare of a certain community. Guo and Saxton (2014) identified a list of advocacy strategies adopted by organizations, and by knowing how these organizations strategized posting behavior, the effectiveness of the messages has the potential to be increased. Besides, the three foci classified in the current study are insufficient to fully account for the whole spectrum of organizational objectives. Capacity building is an overarching term that encompasses anything related to promoting organizations' ability and capability, no matter whether it is tangible (e.g., financial assistance) or intangible (e.g., community outreach), while the other two (prevention and patient advocacy) are very narrowly defined, so it turned out that 63% of the messages were categorized as capacity-building messages. Therefore, by further parsing out capacity building messages, message effectiveness regarding organizations' effort and efficacy on goal fulfillment will be more fully defined.

    Future research should also consider incorporating message-tailoring mechanisms to examine whether or not these organizations compose their messages effectively. Tailoring is a process whereby messages are constructed based upon receivers' individual needs, such as their interests and the context in which the communication takes place, and by identifying message receivers' characteristics, the organizations will be more likely to obtain the desired efficacy (Lustria, Cortese, Noar, & Glueckauf, 2009). Message tailoring has been applied extensively in the field of health communication; it is regarded as effective and essential for disease management, control, and intervention (Bulger & Smith, 1999; Hawkins, Kreuter, Resnicow, Fishbein, & Dijkstra, 2008). Therefore, when delivering messages on social media, health promotion organizations should firstly know with whom they are communicating and then strategically decide what to include in a message.

    Another issue that should be acknowledged is that this study has employed the number of comments as an indicator of the level of engagement with stakeholders. However, users' comments may not be always positive or in agreement with the organization's message. Consequently, future research should delve into the sentiment expressed in these audience comments.

    The current study is at an exploratory stage, and further examination of organizations' efforts in building capacity, the composition of followers, and tailoring tactics to elevate message efficacy are needed. Although some may argue that users' online engagement may not directly link to offline participation (e.g., giving a like to a post encouraging getting a test does not guarantee the action in reality), it has been proved that online social networking could lead to concrete benefits for organizations (Fischer & Reuber, 2011). In other words, social media can actually help an organization to mobilize the resources embedded in its online network in support of organizational endeavors. In brief, examining HIV/AIDS nonprofit organizations' ability when it comes to taking advantage of social media platforms like Facebook is the necessary first step for exploiting the value of social media for disease prevention and intervention processes.

**Table 1.** HIV/AIDS Nonprofit Organizations with Facebook Accounts (N = 110)

| | | | |
|---|---|---|---|
| 1. | AIDS Healthcare Foundation | 56. | HIV Resource Consortium |
| 2. | Lifelong AIDS Alliance | 57. | Alamo Area Resource Center |
| 3. | HIV-AIDS Alliance for Region Two | 58. | AIDS Project of The East Bay |
| 4. | San Francisco AIDS Foundation | 59. | Frannie Peabody Center |
| 5. | Gay Men's Health Crisis | 60. | St. Louis Effort for AIDS |
| 6. | No AIDS Task Force | 61. | Aid for AIDS of Nevada |
| 7. | Desert AIDS Project | 62. | Triangle Area Network |
| 8. | Whitman-Walker Clinic | 63. | Foothill AIDS Project |
| 9. | AIDS Resource Center of Wisconsin | 64. | Delaware HIV Services |
| 10. | AIDS Project Los Angeles | 65. | Greater Ouachita Coalition Providing AIDS Resource & Ed |
| 11. | AIDS ARMS INC | 66. | AIDSNET |
| 12. | AIDS Action Committee of Massachusetts | 67. | AIDS Service Association of Pinellas |
| 13. | Nashville Cares | 68. | Northeast Florida AIDS Network |
| 14. | AIDS Care Group | 69. | Alliance of AIDS Services-Carolina |
| 15. | National Alliance of State and Territorial AIDS Directors | 70. | American Foundation for Children with AIDS |
| 16. | AIDS United | 71. | The Pacific Pride Foundation |
| 17. | AIDS Service Center of Lower Manhattan | 72. | Big Bend Cares |
| 18. | Elton John AIDS Foundation | 73. | South Jersey Against AIDS Alliance |
| 19. | AIDS Resource Center Ohio | 74. | Alder Health Services |
| 20. | Montgomery AIDS Outreach | 75. | AIDS Research Alliance of America |
| 21. | Acadian Concern for AIDS Relief Education and Support | 76. | Heartland Cares |
| 22. | Southern Arizona AIDS Foundation | 77. | Latino Community Services |
| 23. | AID Atlanta | 78. | Health Horizons of East Texas |
| 24. | Valley AIDS Council | 79. | Fredericksburg Area HIV AIDS Support Services |
| 25. | Long Island Association for AIDS Care | 80. | AIDS Network |
| 26. | AIDS Council of Northeastern New York | 81. | Minority AIDS Project |
| 27. | African Comprehensive HIV-AIDS Partnerships | 82. | Evergreen Wellness Advocates |
| 28. | Cascade AIDS Project | 83. | AIDS Alabama |
| 29. | AIDS Services of Austin | 84. | Edge Alliance |
| 30. | Health Services Center | 85. | Community AIDS Resource & Education Services of Southwest Michigan |
| 31. | Resource Center of Dallas | 86. | Open Arms |
| 32. | Interfaith Residence | 87. | AIDS Taskforce of Greater Cleveland |
| 33. | Iris House a Center for Women Living With HIV | 88. | Maui AIDS Foundation |
| 34. | AIDS Foundation Houston | 89. | Lowcountry AIDS Services |
| 35. | Comprehensive AIDS Program of Palm | 90. | Southwest Louisiana AIDS Council |
|     | Beach County                                   |      |                                                                |
| 36. | AIDS Community Resources                       | 91.  | Project Inform                                                 |
| 37. | SAVE Foundation                                | 92.  | Basic NWFL                                                     |
| 38. | Housing Works Services II                      | 93.  | AIDS Volunteers                                                |
| 39. | AIDS Interfaith Residential Services           | 94.  | Nebraska AIDS Project                                          |
| 40. | National Minority AIDS Council                 | 95.  | Palmetto AIDS Life Support Services                            |
| 41. | AIDS Resource Foundation for Children          | 96.  | HIV Education and Prevention Project of Alameda County         |
| 42. | A.H. of Monroe County (AIDS Help)              | 97.  | Oklahoma AIDS Care Fund                                        |
| 43. | Minnesota AIDS Project                         | 98.  | AIDS Interfaith Network                                        |
| 44. | Catawba Care                                   | 99.  | New York City AIDS Memorial                                    |
| 45. | AIDS Connecticut                               | 100. | Black Coalition on AIDS                                        |
| 46. | San Antonio AIDS Foundation                    | 101. | Care for AIDS                                                  |
| 47. | AIDS Project of The Ozarks                     | 102. | Good Samaritan Project                                         |
| 48. | Southern Tier AIDS Program                     | 103. | Boulder County AIDS Project                                    |
| 49. | North Jersey AIDS Alliance                     | 104. | Rural AIDS Action Network                                      |
| 50. | Washington Area Consortium on HIV Infection in Youth | 105. | AIDS Project Greater Danbury                             |
| 51. | New Mexico AIDS Services                       | 106. | Children's AIDS Fund                                           |
| 52. | Damien Center                                  | 107. | Multicultural AIDS Coalition                                   |
| 53. | AIDS Care Ocean State                          | 108. | AIDS Legal Referral Panel of The San Francisco Bay Area        |
| 54. | Pittsburgh AIDS Task Force                     | 109. | Regional AIDS Intercommunity Network of Oklahoma               |
| 55. | Pierce County AIDS Foundation                  | 110. | AIDS Project New Haven                                         |

**Table 2**. Descriptive Statistics for Status Mission/Communicative Functions and Associated Public 'Reaction' (Likes, Comments, & Shares)

| Type of Message | n | % | # of Likes | | | | # of Comments | | | | # of Shares | | | |
|---|---|---|---|---|---|---|---|---|---|---|---|---|---|---|
| | | | M | SD | Min | Max | M | SD | Min | Max | M | SD | Min | Max |
| **All Messages** | 1500 | 100 | 30.59 | 355.34 | 0 | 11004 | 0.59 | 2.62 | 0 | 59 | 3.11 | 14.95 | 0 | 219 |
| **Mission** | | | | | | | | | | | | | | |
| Prevention | 203 | 13.5 | 22.51 | 64.87 | 0 | 501 | 0.10 | 4.90 | 0 | 59 | 4.93 | 17.47 | 0 | 127 |
| Patient advocacy | 148 | 9.9 | 76.23 | 654.15 | 0 | 7,900 | 0.71 | 2.95 | 0 | 29 | 5.78 | 23.02 | 0 | 219 |
| Building capacity | 946 | 63.1 | 29.76 | 363.58 | 0 | 11,004 | 0.59 | 2.01 | 0 | 41 | 2.71 | 13.93 | 0 | 214 |
| Others | 203 | 13.5 | 9.27 | 30.29 | 0 | 387 | 0.37 | 1.30 | 0 | 13 | 1.21 | 6.89 | 0 | 76 |
| **Information** | 693 | 46.2 | 30.75 | 311.58 | 0 | 7,900 | 0.70 | 3.57 | 0 | 59 | 3.82 | 18.41 | 0 | 219 |
| HIV/AIDS info/news | 154 | 10.3 | 16.57 | 58.71 | 0 | 501 | 0.67 | 2.70 | 0 | 23 | 6.06 | 26.62 | 0 | 214 |
| Awareness | 39 | 2.6 | 35.62 | 85.21 | 0 | 408 | 0.72 | 1.57 | 0 | 7 | 6.74 | 17.61 | 0 | 78 |
| Complementary support | 41 | 2.7 | 9.44 | 17.11 | 0 | 86 | 0.20 | 0.60 | 0 | 3 | 1.76 | 4.77 | 0 | 21 |
| News/announcement | 143 | 9.5 | 82.87 | 672.64 | 0 | 7,900 | 0.86 | 4.38 | 0 | 41 | 4.34 | 18.36 | 0 | 142 |
| Medication | 25 | 1.7 | 107.24 | 225.24 | 0 | 823 | 4.72 | 14.42 | 0 | 59 | 26.16 | 55.07 | 0 | 219 |
| Event info | 116 | 7.7 | 4.18 | 6.05 | 0 | 44 | 0.19 | 0.53 | 0 | 3 | 0.45 | 1.25 | 0 | 11 |
| Event update | 92 | 6.1 | 21.85 | 65.84 | 0 | 621 | 0.66 | 1.80 | 0 | 16 | 0.62 | 2.93 | 0 | 25 |
| Cover/profile photo | 22 | 1.5 | 8.23 | 13.44 | 0 | 49 | 0.14 | 0.47 | 0 | 2 | 3.27 | 1058 | 0 | 49 |
| Other | 59 | 3.9 | 10.58 | 42.41 | 0 | 315 | 0.59 | 2.72 | 0 | 20 | 3.07 | 18.42 | 0 | 140 |
| **Community** | 266 | 17.7 | 24.24 | 52.33 | 0 | 517 | 0.75 | 1.67 | 0 | 16 | 3.44 | 17.03 | 0 | 213 |
| Recognition | 198 | 13.2 | 21.46 | 31.63 | 0 | 210 | 0.78 | 1.67 | 0 | 16 | 1.83 | 5.52 | 0 | 46 |
| Holiday | 17 | 1.1 | 10.65 | 13.40 | 0 | 44 | 0.47 | 1.18 | 0 | 4 | 0.18 | 0.39 | 0 | 1 |
| AIDS-related day | 31 | 2.1 | 34.65 | 97.59 | 0 | 517 | 0.39 | 1.82 | 0 | 10 | 8.74 | 28.52 | 0 | 143 |
| Dialogue | 20 | 1.3 | 24.30 | 51.47 | 0 | 228 | 0.90 | 1.41 | 0 | 6 | 3.10 | 6.88 | 0 | 27 |
| **Action** | 541 | 36.1 | 33.51 | 474.06 | 0 | 11,004 | 0.36 | 1.17 | 0 | 13 | 2.04 | 6.48 | 0 | 64 |
| Get tested | 81 | 5.4 | 10.05 | 25.75 | 0 | 198 | 0.21 | 0.94 | 0 | 8 | 1.33 | 3.48 | 0 | 21 |
| Media action | 15 | 1.0 | 18.13 | 51.53 | 0 | 204 | 0.20 | 0.41 | 0 | 1 | 0.33 | 0.49 | 0 | 1 |
| Viewing action | 73 | 4.9 | 11.3 | 22.00 | 0 | 120 | 0.49 | 1.84 | 0 | 13 | 2.55 | 7.81 | 0 | 45 |
| Event promotion | 276 | 18.4 | 12.10 | 32.86 | 0 | 372 | 0.36 | 0.94 | 0 | 7 | 1.96 | 9.54 | 0 | 64 |
| Volunteer | 13 | 0.9 | 9.62 | 18.15 | 0 | 68 | 0.08 | 0.28 | 0 | 1 | 2.23 | 3.88 | 0 | 14 |
| Donation | 85 | 5.7 | 145.44 | 1,192.77 | 0 | 11,004 | 0.38 | 1.20 | 0 | 7 | 2.28 | 5.48 | 0 | 29 |

Table 3. Negative Binomial Regressions with Dependent Variables: # of Likes, Comments, and Shares

| | # Likes | # Comments | # Shares | # Likes | # Comments | # Shares |
|---|---|---|---|---|---|---|
| *Mission Focus* | | | | | | |
| Prevention | 0.73** | 1.27*** | 0.90** | 1.04*** | 1.31** | 1.12** |
| | (0.34) | (0.47) | (0.37) | (0.36) | (0.54) | (0.46) |
| Patient advocacy | 0.56 | 1.02** | 0.23 | 0.69* | 1.32** | 0.36 |
| | (0.36) | (0.49) | (0.41) | (0.37) | (0.54) | (0.49) |
| Building capacity | 0.73*** | 0.87** | 0.48 | 0.78*** | 1.06** | 0.62* |
| | (0.27) | (0.39) | (0.31) | (0.26) | (0.42) | (0.36) |
| *ICA Framework* | | | | | | |
| Community | 0.16 | -0.35 | 0.2 | | | |
| | (0.24) | (0.32) | (0.24) | | | |
| Action | -0.60*** | -0.80*** | -0.78*** | | | |
| | (0.2) | (0.27) | (0.23) | | | |
| *Information sub-codes* | | | | | | |
| HIV info/news | | | | -0.19 | -0.09 | -0.44 |
| | | | | (0.47) | (0.67) | (0.58) |
| Awareness | | | | -0.69 | -1.14 | -1.12 |
| | | | | (0.6) | (0.85) | (0.75) |
| Complementary support | | | | -0.62 | -1.49* | -1.28 |
| | | | | (0.62) | (0.90) | (0.92) |
| Organizational news | | | | 0.69* | 0.1 | -0.15 |
| | | | | (0.42) | (0.6) | (0.52) |
| Medication | | | | 1.11* | 0.94 | 0.47 |
| | | | | (0.67) | (0.89) | (0.73) |
| Event info | | | | -1.15** | -2.20*** | -1.48** |
| | | | | (0.46) | (0.76) | (0.65) |
| Event update | | | | 0.39 | -1.87** | -0.26 |
| | | | | (0.48) | (0.77) | (0.59) |
| Cover/profile photo | | | | -0.38 | 0.06 | -1.56 |
| | | | | (0.65) | (0.89) | (1.19) |
| *Community sub-codes* | | | | | | |
| Recognition | | | | 0.28 | -0.98 | -0.19 |
| | | | | (0.44) | (0.66) | (0.55) |
| National holiday/Holiday | | | | 0.37 | -2.38 | -0.07 |
| | | | | (0.7) | (1.82) | (0.85) |
| AIDS-related day | | | | 0.06 | -0.3 | -1.71* |
| | | | | (0.61) | (0.84) | (0.96) |
| Dialogue | | | | -0.47 | -1.34 | -0.10 |
| | | | | (0.7) | (1.04) | (0.78) |
| *Action sub-codes* | | | | | | |
| Get tested | | | | -1.16** | -1.73** | -2.4*** |
|  |  |  |  | (0.56) | (0.82) | (0.82) |
| Media action |  |  |  | -0.02 | -2.51 | -1.45 |
|  |  |  |  | (0.77) | (1.58) | (1.26) |
| Viewing action |  |  |  | -0.57 | -1.05 | -0.84 |
|  |  |  |  | (0.51) | (0.75) | (0.66) |
| Event promotion |  |  |  | -0.46 | -1.13$^*$ | -1.10$^{**}$ |
|  |  |  |  | (0.43) | (0.64) | (0.56) |
| Volunteer |  |  |  | -0.57 | -0.77 | -2.45 |
|  |  |  |  | (0.82) | (1.14) | (1.93) |
| Donation |  |  |  | -0.37 | -1.04 | -1.21$^*$ |
|  |  |  |  | (0.50) | (0.73) | (0.66) |
| Hashtag count | -0.11 | -0.07 | -0.20$^*$ | -0.10 | -0.07 | -0.18$^*$ |
|  | (0.08) | (0.11) | (0.11) | (0.06) | (0.09) | (0.09) |
| *Organizational Controls* |  |  |  |  |  |  |
| # of Followers (1,000s) | -0.0023 | -0.0200$^{***}$ | 0.0029 | -0.0044 | -0.0153$^{**}$ | 0.0018 |
|  | (0.0058) | (0.0073) | (0.0063) | (0.0045) | (0.0061) | (0.0057) |
| Assets (10,000s) | 0.0003 | 0.008$^{***}$ | 0.0001 | 0.0003$^{**}$ | 0.0006$^{***}$ | 0.0001 |
|  | (0.002) | (0.0002) | (0.0002) | (0.0001) | (0.0002) | (0.0002) |
| Constant | 2.05$^{***}$ | -0.04 | -1.03$^{***}$ | 1.87$^{***}$ | 0.26 | -0.73$^*$ |
|  | (0.24) | (0.36) | (0.28) | (0.34) | (0.48) | (0.41) |
| Model Significance ($X^2$) | 14661$^{***}$ | 19100$^{***}$ | 5981.9$^{***}$ | 8602.1$^{***}$ | 11990$^{***}$ | 4701.9$^{***}$ |
| Log Likelihood | -5539.8 | -2967.1 | -1384 | -5398.4 | -276739 | -1332.5 |
| N | 1500 | 1500 | 1500 | 1500 | 1500 | 1500 |

Note. $^*p < 0.10$, $^{**}p < 0.05$, $^{***}p < 0.01$; standard errors in parentheses; for the ICA framework tested in the first three regressions, Informational messages constitute the omitted or baseline category, the category against which the Community and Action categories can be compared. For ICA sub-categories (regressions 4-6), *Other* messages were omitted.

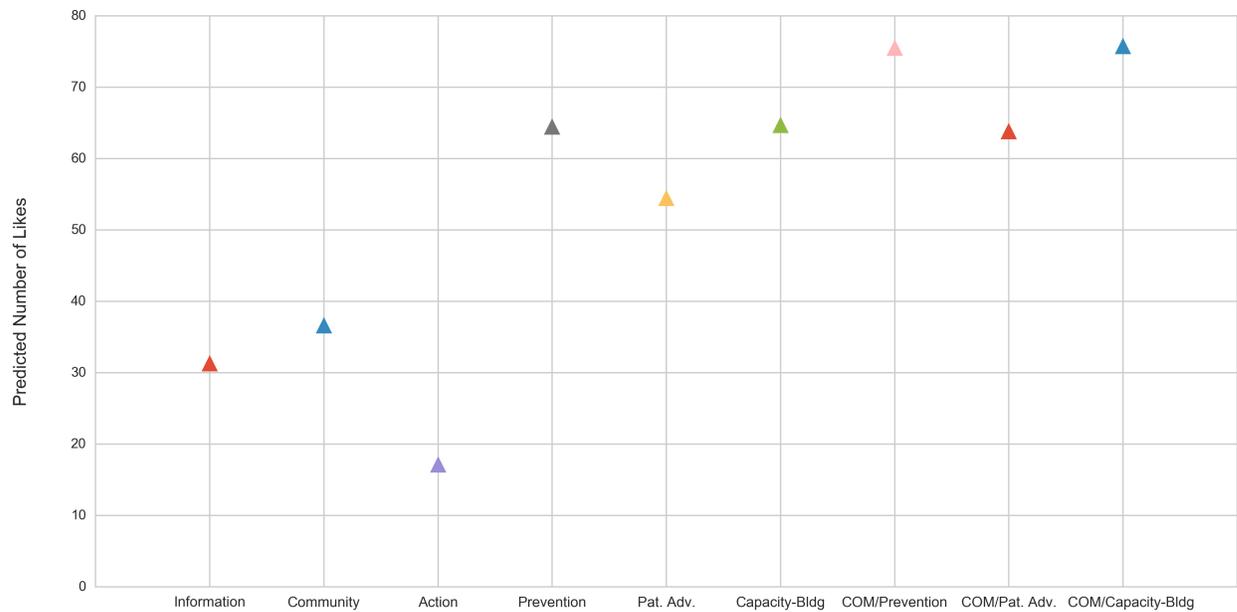

**Figure 1. Predicted Number of Likes for Various Configurations of Message Types**

*Note:* Figure shows predicted number of likes based on post-regression estimations based on model 1 in Table 3. The first three triangles show the predicted count for an *information, community,* and *action* message, respectively, for a non-mission-related message sent by an organization with an average number of followers and average asset size. The fourth through sixth triangles show the predicted number of likes for informational messages that are mission-focused on, respectively, *Prevention, Patient Advocacy,* and *Capacity-Building*. The final three triangles are for messages that combine a *Community* orientation with the three respective types of mission focus. In all cases predictions are for messages sent by organizations with average values for assets and number of followers.

**Appendix 1.** *Descriptions of Each Mission Focus and Sub-categories of ICA Framework*

| Category and Definition | Example |
|---|---|
| **I. MISSION** | |
| **Prevention** – Information about preventive measures or practices intended to constrain the spread of HIV (e.g., the CDC's Pre-exposure prophylaxis (PrEP) program). | *NASTAD*: Wondering how health departments are implementing PrEP to end HIV? Check out our latest post on the topic: http://bit.ly/WBDz7c #HIV #AIDS #PrEP |
| **Patient advocacy** – Information containing resources (e.g., financial assistance, social support groups) that could help improve the quality of life for people living with HIV. | *Whitman-Walker Health*: Are the holidays a difficult time for you? At Whitman-Walker, we offer one-on-one and group peer support. Check out www.whitman-walker.org/peersupport |
| **Building capacity** – Dissemination of facts and knowledge not related to prevention or patient advocacy but which are needed to fulfill organizational goals, including advertising events and donation requests. Also collaboration with the community or other organizations. | *NO/AIDS Task Force:* It's not too late to buy your tickets for the Chevron #ArtAgainstAIDS gala tomorrow ... Don't miss out! |
| **Other** - Statuses that are not mission-related. | *HIV/AIDS Alliance for Region Two, Inc. (HAART)*: From the HAART family to your family Merry Christmas! |
| **II. I-C-A FRAMEWORK** | |
| *Information* | |
| - **HIV/AIDS info/news** - Information that purely shares facts, knowledge, or statistical reports about HIV. | *GMHC*: New York is number nine on the list of the 25 cities with the highest rates of HIV infection. bit.ly/1uW3gfy |
| - **Awareness** - Reminders of preventive behavior (e.g., safer sex) or attempts to make audiences more conscious about the importance/severity of HIV-related situations. | *Catawba Care*: Happy Valentines Day! Protect your love...use a condom! |
| - **Complementary support** - Information to make the lives of people living with HIV more financially affordable (e.g., housing, insurance plan) or provide venues for social support (e.g., support groups). | *AIDS Care Group*: Our Positively Recovering Group is meeting today in Chester. Join them to talk about your path after discovering your HIV status. |
| - **Organizational News/Announcement** - Information about recent achievements or organizational updates (e.g., recruitment). Additionally, the dissemination of information about accessible resources for those who are in need (e.g., introduction of a physician). | *Desert AIDS Project*: Our bright new bike rack just installed on Friday. Now we just have to wait for it to get cool enough to ride during the day again! |
| - **Medication** - Information about treatment medication and preventive drugs, including announcements about new drugs or issues surrounding controversial medicine. | *AIDS Resource Center of Wisconsin – ARCW*: Interesting article about an experimental new drug that could potentially protect people from the spread of the AIDS Virus. |
| - **Event info** - Information regarding details of an event, such as time, date, place, or direct links to an event. | *Triangle AIDS Network*: To order tickets to the Garden Party Please use this link : http://www.tanbmt.com/tan-event-tickets/ |

- **Event update** - Any change of an event (e.g., cancellation/postponing) or recapitulation of an event (e.g., photos.)
- **Cover/Profile photo** - Change of an organization's Facebook cover or profile photo.
- **Other** - Posts that did not fall under any of the categories above.

*Community*
- **Recognition** - Showing appreciation to individual(s) for their dedication/donation to an event or the organization. It can also be a descriptive acknowledgment of a personal achievement or contribution to the organization or an HIV-related cause/event.
- **National holiday/Holiday** - Recognition of national holidays (e.g., Thanksgiving)
- **AIDS-related day** - A day that is set to raise awareness of and fight against the disease.
- **Dialogue** - The status encourages the audience to respond with their feedback or opinions.

*Action*
- **Get tested** - Encouraging/asking their followers to undergo HIV/STD testing, including the use of hashtags (e.g., #GetTested).
- **Media action** - The audience is asked to "share" or "like" the post on social media.
- **Viewing action** - The messages utilized verbs such as "learn," "read," or "watch" to ask the audience to read articles, see photos, or watch a video.
- **Event promotion** – Posts are to encourage participation in an event.
- **Volunteer** - Recruitment of volunteers for organizational events or activities.
- **Donation** - Posts requesting donations, whether in the form of money or goods.

*Resource Center*: Due to weather, the Center's United Black Ellument Interfaith Panel Discussion is CANCELED and will be rescheduled.

*Long Island Association for AIDS Care – liaac*: Stay safe. Keep learning.

*AIDS Action Committee*: Elton John, who recently gave Fenway and the HRC a $300,000 grant for HIV/AIDS education and care, on why the fight against AIDS isn't over.

*Doorways Interfaith AIDS Housing and Services*: Merry Christmas from all of us at Doorways!

*Cascade AIDS Project*: We at CAP are getting so excited about World AIDS Day. What events will you be attending this year?

*AVOL (AIDS Volunteers, Inc.)*: Happy Dining Out For Life Day, Lexington! Tell us in the comment section where you will be Dining Out For Life!

*Minnesota AIDS Project*: Knowing your status is the first step in preventing the spread of HIV. #GetTested #StopHIV http://ow.ly/u2ncv

*AVOL (AIDS Volunteers, Inc.)*: It's Day 8 of our educational campaign, #GiveGreatEd. Have you shared any of our images yet?

*Frannie Peabody Center*: Learn more about The AIDS Generation: Stories of Survival and Resilience

*AIDS Action Committee*: Just 12 days until AIDS Walk Boston! Sign up today and help us continue to care for and empower those at risk and living with HIV/AIDS.

*Metro TeenAIDS*: Need community service hours? Come by @ 5pm today for our #volunteer orientation for 13-24 y/o at 651 Penn Ave. SE.

*Pittsburgh AIDS Task Force*: Have a percentage of your donation to PATF matched by the Pittsburgh Foundation on Tuesday, May 6.